\input{aipcheck}

\documentclass[
    ,final            
  ]
  {aipproc}

\usepackage{amssymb}
\usepackage{amsmath}

\layoutstyle{8x11double}

\newcommand{\eVq}{\mathrm{eV}^2}
\newcommand{\Dmq}{\Delta m^2}
\newcommand{\stheta}{\sin^22\theta_{13}}

\begin{document}

\title{
\flushright\mbox{\normalsize CERN-TH-2007-207}\\[10mm]
\centering
Neutrino oscillations: present status and outlook
\footnote{Plenarly talk at NuFact07, Okayama, Japan.}
}

\classification{14.60.Lm, 14.60.Pq, 14.60.St}
\keywords      {Neutrino oscillations}

\author{Thomas Schwetz}{
  address={Physics Department, Theory Division, CERN, CH--1211 Geneva
  23, Switzerland} }

\begin{abstract}
I summarize the status of three-flavour neutrino oscillations with
date of Oct.\ 2007, and provide an outlook for the developments to be
expected in the near future. Furthermore, I discuss the status of
sterile neutrino oscillation interpretations of the LSND anomaly in
the light of recent MiniBooNE results, and comment on implications for
the future neutrino oscillation program.
\end{abstract}

\maketitle



Thanks to the spectacular developments in neutrino oscillation
experiments in the last years we have now a rough picture of the
parameters governing three-flavour oscillations (see also
Ref.~\cite{Kajita} for an overview): There are two mass-squared
differences separated roughly by a factor 30, there are two large
mixing angles ($\theta_{23}$, which could even be $45^\circ$, and
$\theta_{12}$, which is large but smaller than $45^\circ$ at very high
significance), and one mixing angle which has to be small
($\theta_{13}$). Present data is consistent with two possibilities for the
neutrino mass ordering, conventionally parametrized by the sign of
$\Dmq_{31}$: In the normal ordering ($\Dmq_{31} > 0$) the mass state
which contains predominantly the electron neutrino has the smallest
mass, whereas in the inverted ordering ($\Dmq_{31} < 0$) it is part of
a nearly degenerate doublet of mass states which is separated from the
lightest neutrino mass by $|\Dmq_{31}|$.

\section{Global three-flavour analysis}

In this section I present an update on the determination of
three-neutrino oscillation parameters from a global analysis of latest
world neutrino oscillation data from solar, atmospheric, reactor, and
accelerator experiments.
These results are based on work in collaboration with M.~Maltoni,
M.~Tortola and J.W.F.~Valle, published in
Refs.~\cite{Maltoni:2003da,Maltoni:2004ei} (see also the arXiv
version~6 of Ref.~\cite{Maltoni:2004ei} for updated results).  The
present determination of the three-flavour oscillation parameters is
summarized in Tab.~\ref{tab:summary}, where the best fit points and
the $2\sigma$ and $3\sigma$ allowed ranges are given.

\begin{table}[t] \centering
    \catcode`?=\active \def?{\hphantom{0}}
    \begin{tabular}{lccc}
        \hline
        \tablehead{1}{l}{b}{Parameter} & 
        \tablehead{1}{c}{b}{Best fit} & 
        \tablehead{1}{c}{b}{2$\sigma$} & 
        \tablehead{1}{c}{b}{3$\sigma$} 
        \\
        \hline
        $\Delta m^2_{21}\:\: [10^{-5}\eVq]$
        & 7.6?? & 7.3--8.1 & 7.1--8.3 \\
        $|\Delta m^2_{31}|\: [10^{-3}\eVq]$
        & 2.4?? & 2.1--2.7 & 2.0--2.8 \\
        $\sin^2\theta_{12}$
        & 0.32? & 0.28--0.37 & 0.26--0.40\\
        $\sin^2\theta_{23}$
        & 0.50? & 0.38--0.63 & 0.34--0.67\\
        $\sin^2\theta_{13}$
        & 0.007 &  $\leq$ 0.033 & $\leq$ 0.050 \\
        \hline
    \end{tabular}
\caption{ \label{tab:summary} 
  Best-fit values, 2$\sigma$ and 3$\sigma$ intervals (1 d.o.f.) for
  the three--flavour neutrino oscillation parameters from global data
  including solar, atmospheric, reactor (KamLAND and CHOOZ) and
  accelerator (K2K and MINOS) experiments.}
\end{table}

\begin{figure}
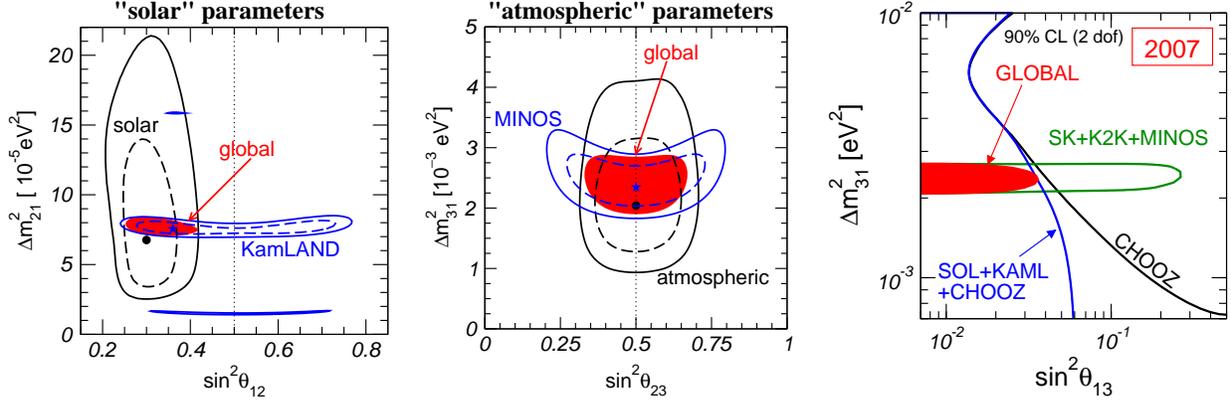

  \includegraphics[height=5.3cm]{figs/sol_vs_kaml-2007.eps}
  \quad
  \includegraphics[height=5.3cm]{figs/atm_vs_lbl-2007.eps}
  \quad
  \includegraphics[height=5.3cm]{figs/th13-2007.eps} 
  \caption{Determination of the leading ``solar'' (left) and
  ``atmospheric'' (middle) oscillation parameters from the interplay
  of data from artificial and natural neutrino sources. Right:
  Constraint on $\theta_{13}$ from the global analysis of neutrino
  data.}  \label{fig:status}
\end{figure}

This analysis includes new data released this summer by the
MINOS~\cite{Collaboration:2007zz, minos-talk} and
KamLAND~\cite{KamLAND:2007, Araki:2004mb} collaborations which lead to
an improved determination of the mass-squared differences $|\Delta
m^2_{31}|$ and $\Delta m^2_{21}$, respectively, mainly due to the
precise spectral information.
New MINOS data have been collected from June 2006 to July 2007
(Run-IIa), and they have been analyzed together with the first data
sample (Run-I), with a total exposure of 2.5$\times$10$^{20}$~p.o.t.
In total, 563 $\nu_\mu$ events have been observed at the far detector,
while 738$\pm$30 events were expected for no oscillation. 
KamLAND data presented at TAUP2007~\cite{KamLAND:2007} correspond to a
total exposure of 2881 ton$\cdot$year, almost 4 times larger than 2004
data~\cite{Araki:2004mb}.  Apart from the increased statistics also
systematic uncertainties have been improved: Thanks to the full volume
calibration the error on the fiducial mass has been reduced from 4.7\%
to 1.8\%. The main limitation for the $\Delta m^2_{21}$ measurement
comes now from the uncertainty on the energy scale of 1.5\%.  Details
of our KamLAND analysis are described in appendix~A of
Ref.~\cite{Maltoni:2004ei}.  We use the data binned in equal bins in
$1/E$ to make optimal use of spectral information, we take into
account the (small) matter effect and carefully include various
systematics. As previously we restrict the analysis to the prompt
energy range above 2.6~MeV to avoid large contributions from
geo-neutrinos and backgrounds. In that energy range 1549 reactor
neutrinos events and a background of 63 events are expected without
oscillations, whereas the observed number of events is 985.

Fig.~\ref{fig:status} illustrates how the determination of the
leading ``solar'' ($\theta_{12}$ and $\Dmq_{21}$) and ``atmospheric''
($\theta_{23}$ and $|\Dmq_{31}|$) oscillation parameters emerges from
the complementarity of data from natural (sun and atmosphere) and
men-made (reactor and accelerator) neutrino sources.  Spectral
information from KamLAND data leads to an accurate determination of
$\Delta m^2_{21}$ with the remarkable precision of 5\% at
$2\sigma$. KamLAND data start also to contribute to the lower bound
on $\sin^2\theta_{12}$, whereas the upper bound is still dominated by
solar data, most importantly by the CC/NC solar neutrino rate measured
by SNO~\cite{Aharmim:2005gt}. Moreover, as evident from
Fig.~\ref{fig:status} solar data fixes the octant of $\theta_{12}$,
thanks to the MSW mechanism~\cite{Wolfenstein:1978ue, Mikheev:1985gs}
due to matter effects inside the sun, whereas the small matter effect
in KamLAND cannot break the symmetry between the first and second
$\theta_{12}$ octants.

We find a similar complementarity also in the determination of the
atmospheric oscillation parameters, see middle panel in
Fig.~\ref{fig:status}. In this case the $|\Delta m^2_{31}|$
determination is dominated by data from the MINOS long-baseline
$\nu_\mu$ disappearance experiment, which by now largely supersedes
the pioneering K2K measurement~\cite{Aliu:2004sq}, although in the
global analysis the latter still contributes slightly to the lower
bound on $|\Delta m^2_{31}|$. The determination of the mixing angle
$\theta_{23}$ is dominated by atmospheric neutrino data from
Super-Kamiokande~\cite{Ashie:2005ik}, leading to a best fit point at
maximal mixing.\footnote{Small deviations from maximal mixing due to
sub-leading three-flavour effects are discussed in
Refs.~\cite{Fogli:2006qg, GonzalezGarcia:2007ib}. At present such
deviations are not statistically significant.} The sign of $\Dmq_{31}$
(i.e., the neutrino mass hierarchy) is undetermined by present data.

Similar to the case of the leading oscillation parameters, also the
bound on $\theta_{13}$ emerges from an interplay of different data
sets, as illustrated in right panel of Fig.~\ref{fig:status}. An
important contribution to the bound comes, of course, from the CHOOZ
reactor experiment~\cite{Apollonio:2002gd} combined with the
determination of $|\Dmq_{31}|$ from atmospheric and long-baseline
experiments. However, due to a complementarity of low and high energy
solar data, as well as solar and KamLAND data also solar+KamLAND
provide a non-trivial constraint on $\theta_{13}$, see e.g.,
\cite{Maltoni:2003da,Maltoni:2004ei,Goswami:2004cn}.  We obtain
at 90\%~CL ($3\sigma$) the following limits:
\[
  \sin^2\theta_{13} \le \left\lbrace \begin{array}{l@{\qquad}l}
      0.051~(0.084) & \mbox{solar+KamLAND} \\
      0.028~(0.059) & \mbox{CHOOZ+atm+LBL} \\
      0.028~(0.050) & \mbox{global data}
    \end{array} \right.
\]
In the global analysis we find a slight weakening of the upper bound
on $\sin^2\theta_{13}$ due to the new data from 0.04 (see
Ref.~\cite{Schwetz:2006dh} or v5 of \cite{Maltoni:2004ei}) to 0.05 at
$3\sigma$. The reason for this is two-fold. First, the shift of the
allowed range for $|\Delta m^2_{31}|$ to lower values due to the new
MINOS data implies a slightly weaker constraint on $\sin^2\theta_{13}$
(cf.\ Fig.~\ref{fig:status}), and second, the combination of solar and
new KamLAND data prefers a slightly non-zero value of
$\sin^2\theta_{13}$ which, though not statistically significant, also
results in a weaker constraint in the global fit. Note also that
sub-leading effects in atmospheric neutrino data have an impact on the
bound on $\theta_{13}$ at that level, as discussed in
Ref.~\cite{Schwetz:2006dh}.

\begin{figure}
  \includegraphics[height=4.2cm]{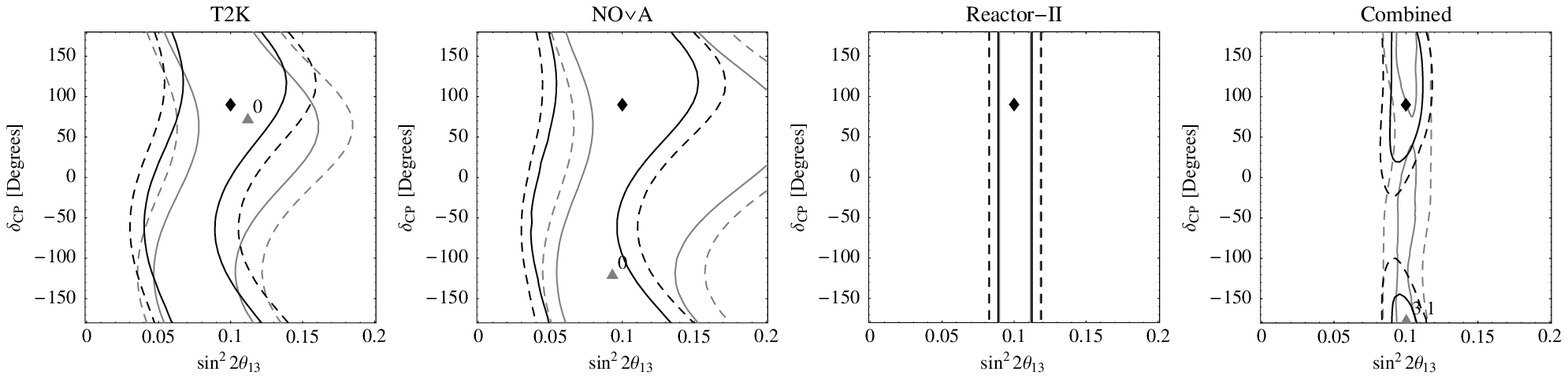}
  \caption{The $90 \%$ CL (solid) and $3 \sigma$ (dashed) allowed
   regions (2 d.o.f.) in the $\stheta$-$\delta$ plane for the true
   values $\stheta=0.1$ and $\delta=90^\circ$ for T2K, NO$\nu$A, a
   reactor experiment, and the combination. The black curves refer to
   the allowed regions for the normal mass hierarchy, whereas the gray
   curves refer to the inverted hierarchy. The best fits are marked as
   diamonds (normal hierarchy) and triangles (inverted hierarchy). For
   the latter, the $\Delta\chi^2$-value with respect to the best-fit
   point is also given~\cite{Huber:2004ug}.}  \label{fig:del-theta}
\end{figure}

\section{Outlook for the near future} %

In the following I try to give some outlook on developments in
neutrino oscillations to be expected at a time scale of 5 to 10
years~\cite{Huber:2004ug}. In this time-frame we expect results from a
new generation of reactor experiments, Double-Chooz~\cite{dchooz} and
Daya-Bay~\cite{dayabay}, as well as the next generation of
long-baseline superbeam experiments T2K~\cite{t2k} and
NO$\nu$A~\cite{nova}.

The currently running MINOS experiment will improve further the
determination of $|\Delta m^2_{31}|$ with accumulating
statistics. Once results on $\nu_\mu$ disappearance become available
from the T2K and/or NO$\nu$A experiments a determination of this
parameter at the level of a few percent at $2\sigma$ will be
obtained~\cite{Huber:2004ug, t2k, nova} (currently 12\%, cf.\
Tab.~\ref{tab:summary}), and also $\sin^2\theta_{23}$ is likely to be
measured with a precision better than present atmospheric neutrino
data.

Certainly the main goal of the upcoming experiments is the
determination of $\theta_{13}$. Reactor experiments aim at this goal
by exploring the disappearance of $\bar\nu_e$. The corresponding
survival probability is given to very good accuracy by
\[
P_{ee} = 1 - \stheta \sin^2\frac{\Delta m^2_{31} L}{4 E_\nu} \,.
\]
This simple dependence shows that reactor experiments provide a clean
measurement of $\stheta$, not affected by correlation or degeneracies
with other unknown parameters~\cite{Huber:2003pm}. The main issue in
such an experiment are statistical and systematical errors, where the
latter are going to be addressed by comparing data for near and far
detectors. In contrast, the superbeam experiments look for the
appearance of $\nu_e$ from a beam consisting initially mainly of
$\nu_\mu$.  At leading order in the small parameters $\sin
2\theta_{13}$ and $\tilde\alpha \equiv
\sin 2\theta_{12} \Dmq_{21} L / 4 E_\nu$ the relevant oscillation
probability (in vacuum, for simplicity) is
\begin{eqnarray}
P_{\mu e} &=&  \stheta \, \sin^2\theta_{23} \, \sin^2\Delta  + 
              \tilde\alpha^2 \, \cos^2\theta_{23}  \nonumber\\
          &+& \sin 2\theta_{13}  \sin 2\theta_{23} \, \tilde\alpha
              \sin\Delta \cos(\Delta \pm \delta) \,,\nonumber
\end{eqnarray}
where $\Delta \equiv \Dmq_{31} L / 4 E_\nu$, and '$+$' ('$-$') holds
for neutrinos (anti-neutrinos). This expression shows that there is a
complicated correlation of $\stheta$ with other parameters, especially
with the CP phase $\delta$. This effect is illustrated in
Fig.~\ref{fig:del-theta}, where the allowed region in the plane of
$\stheta$ and $\delta$ is shown for T2K, NO$\nu$A, a reactor
experiment, and the combination, assuming an input value of $\stheta =
0.1$ and $\delta = 90^\circ$. For the superbeams the allowed regions
show a typical 'S'-shape, reflecting the trigonometric dependence of
the probability on $\delta$. Furthermore, solutions with the wrong
mass hierarchy (gray curves in the figure) introduce another ambiguity
in the interpretation. On the other hand, the figure shows that a
reactor experiment can determine $\stheta$ unambiguously. The right
most panel illustrates the situation which could emerge from the global
analysis of these experiments: A relative good determination of
$\theta_{13}$, some information on $\delta$ (though CP violation
cannot be established), which however is largely corrupted by the
ambiguity in the mass hierarchy, which cannot be resolved in this
particular example ($\Delta\chi^2$ of the wrong hierarchy is only 3.1
in the global analysis).

\begin{figure}
  \includegraphics[height=7cm]{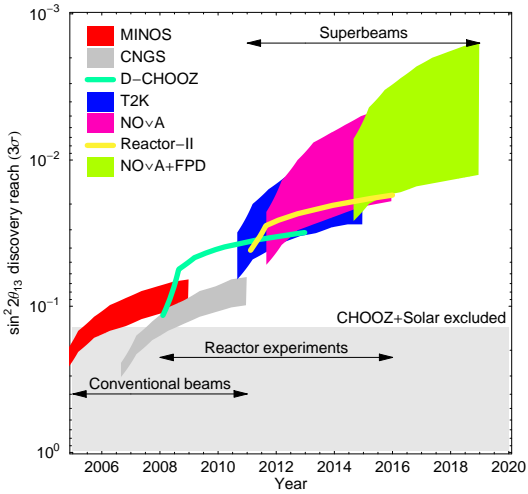}
  \caption{Evolution of the $3\sigma$ discovery potential of a
  non-zero value of $\theta_{13}$ of upcoming
  experiments. Figure~from~\cite{Albrow:2005kw}.}  \label{fig:th13-evol}
\end{figure}

Fig.~\ref{fig:th13-evol} shows the evolution of the $\theta_{13}$ discovery
reach as a function of time, where of course a significant uncertainty
is associated with the horizontal axis. 
The complementarity of beam and reactor experiments is also visible in
that figure: The wide bands for the beam experiments follow from the
impact of the (unknown) true value of $\delta$, which could be in
favor for discovering a non-zero value of $\theta_{13}$ or not. In
contrast, reactor experiments do not depend on the CP-phase and their
reach is just determined by statistics and systematics.

Clearly, with the next generation of experiments we are entering the
era of precision measurements, at the level of 1\%. The are many ideas
on how to go beyond this level, performing high precision measurements
addressing questions like leptonic CP violation or the type of the
neutrino mass hierarchy, among them high-intensity superbeams, beta
beams or neutrino factories. These options have been discussed at this
conference.

\section{LSND and MiniBooNE results}%

\begin{figure}
  \includegraphics[height=7cm]{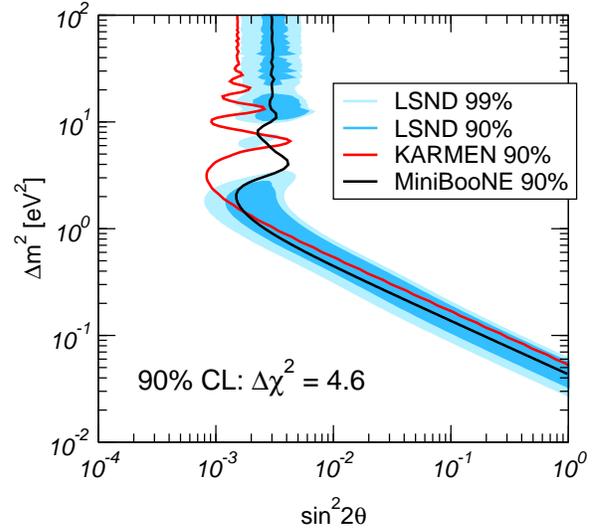}
  \caption{Exclusion contours at 90\% (2~d.o.f.) for MiniBooNE and
  KARMEN compared to the LSND allowed region.}
  \label{fig:LSND}
\end{figure}

Reconciling the LSND evidence~\cite{Aguilar:2001ty} for $\bar\nu_\mu
\to \bar\nu_e$ oscillations with the global neutrino data reporting
evidence and bounds on oscillations remains a long-standing problem
for neutrino phenomenology. Recently the MiniBooNE
experiment~\cite{miniboone, AguilarArevalo:2007it} added more
information to this question. This experiment searches for
$\nu_\mu\to\nu_e$ appearance with a very similar $L/E_\nu$ range as
LSND. No evidence for oscillations is found and the results are
inconsistent with a two-neutrino oscillation interpretation of LSND at
98\%~CL~\cite{AguilarArevalo:2007it}. The exclusion contour from
MiniBooNE is shown in Fig.~\ref{fig:LSND} in comparison to the LSND
allowed region and the previous bound from the KARMEN
experiment~\cite{Armbruster:2002mp}, all in the framework of 2-flavour
oscillations.

The standard ``solution'' to the LSND problem is to introduce one or
more sterile neutrinos at the eV scale in order to provide the
required mass-squared difference to accommodate the LSND signal in
addition to ``solar'' and ``atmospheric'' oscillations.  However, in
such schemes there is a sever tension between the LSND signal and
short-baseline disappearance experiments, most importantly
Bugey~\cite{Declais:1994su} and CDHS~\cite{Dydak:1983zq}, with some
contribution also from atmospheric neutrino
data~\cite{Bilenky:1999ny}.  I report here the results
from~\cite{Maltoni:2007zf}, where a global analysis including the
MiniBooNE results has been performed in schemes with one, two and
three sterile neutrinos (see also~\cite{karagiorgi}).

\begin{figure}
  \includegraphics[height=7cm]{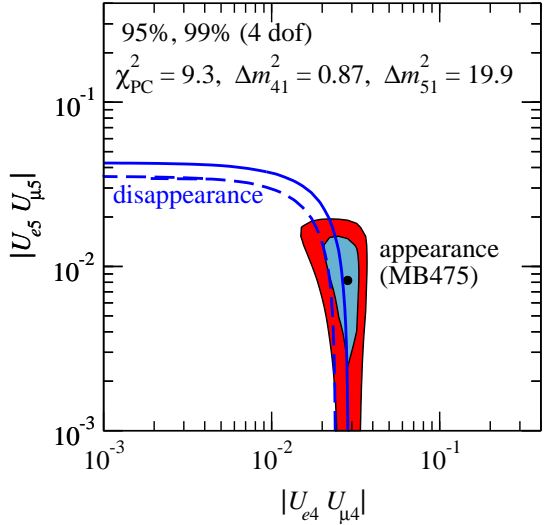}
  \caption{Section of the 4-dimensional volumes allowed at 95\% and
  99\%~CL in the (3+2) scheme from SBL appearance and disappearance
  experiments in the space of the parameters in common to these two
  data sets. The values of $\Dmq_{41}$ and $\Dmq_{51}$ of the
  displayed sections correspond to the point in parameter space where
  the two allowed regions touch each other (at a $\Delta\chi^2 =
  9.3$).}\label{fig:3+2}
\end{figure}

Four-neutrino oscillations within so-called (3+1) schemes have been
only marginally allowed before the recent MiniBooNE results (see,
e.g., Refs.~\cite{Maltoni:2002xd, Maltoni:2004ei, Sorel:2003hf}), and
become even more disfavored with the new data, at the level of
$4\sigma$~\cite{Maltoni:2007zf}.  Five-neutrino oscillations in (3+2)
schemes~\cite{Sorel:2003hf} allow for the possibility of CP violation
in short-baseline oscillations~\cite{Karagiorgi:2006jf}. Using the
fact that in LSND the signal is in anti-neutrinos, whereas present
MiniBooNE data is based on neutrinos, these two experiments become
fully compatible in (3+2) schemes~\cite{Maltoni:2007zf}.
However, in the global analysis the tension between appearance and
disappearance experiments remains unexplained. This problem is
illustrated in Fig.~\ref{fig:3+2} where sections through the allowed
regions in the parameter space for appearance and disappearance
experiments are shown. An opposite trend is clearly visible: while
appearance data require non-zero values for the mixing of $\nu_e$ and
$\nu_\mu$ with the eV-scale mass states 4 and 5 in order to explain
LSND, disappearance data provide an upper bound on these mixing. The
allowed regions touch each other at $\Delta\chi^2 = 9.3$, and a
consistency test between these two data samples yields a probability
of only $0.18\%$, i.e., these models can be considered as disfavoured
at the $3\sigma$ level. Furthermore, when moving from 4 neutrinos to 5
neutrinos the fit improves only by 6.1 units in $\chi^2$ by
introducing 4 more parameters, showing that in (3+2) schemes the
tension in the fit remains a sever problem. This is even true in the
case of three sterile neutrinos, since adding one more neutrino to
(3+2) cannot improve the situation~\cite{Maltoni:2007zf}.\footnote{In
Ref.~\cite{Schwetz:2007cd} I have pointed out that an exotic sterile
neutrino with energy dependent mass or mixing can resolve these
tensions.}

In view of this somewhat ambiguous situation I comment in the
following on the impact of sterile neutrinos (with masses and mixing
relevant for LSND) for future neutrino oscillation experiments. As
discussed in the previous section the typical
$\theta_{13}$-sensitivity of the next generation of experiments
(Double-Chooz, T2K, NO$\nu$A) is $\stheta \gtrsim 1\%$, cf.\
Fig.~\ref{fig:th13-evol}. This should be compared to the size of the
appearance probability observed in LSND: $P_\mathrm{LSND} \approx
0.26\%$. Hence, if $\theta_{13}$ is large enough to be found in those
experiments sterile neutrinos may introduce some sub-leading effect,
but their presence cannot be confused with a non-zero $\theta_{13}$.
Nevertheless, I argue that it could be worth to look for sterile
neutrino effects in the next generation of experiments. They would
introduce (mostly energy averaged) effects, which could be visible as
disappearance signals in the near detectors of these experiments.
This has been discussed in \cite{Bandyopadhyay:2007rj} for the
Double-Chooz experiment, but also the near detectors at superbeam
experiments should be explored.\footnote{An interesting effect of
(3+2) schemes has been pointed out recently for high energy
atmospheric neutrinos in neutrino telescopes \cite{Choubey:2007ji}.}

However, for the subsequent generation of oscillation experiments
aiming at sub-percent level precision to test CP violation and the
neutrino mass hierarchy, the question of LSND sterile neutrinos is
highly relevant \cite{Donini:2001xp,Dighe:2007uf}. They will lead to a
miss-interpretation or (in the best case) to an inconsistency in the
results. If eV scale steriles exist with mixing relevant for LSND the
optimization in terms of baseline and $E_\nu$ of high precision
experiments has to be significantly changed. Therefore, I argue that
it is important to settle this question at high significance before
decisions on high precision oscillation facilities are taken.

\bigskip

\textbf{Acknowledgment.}
  The author acknowledges support from the BENE network.

\bibliographystyle{aipproc}   

\bibliography{nufact07-proc-arxiv}

\IfFileExists{\jobname.bbl}{}
 {\typeout{}
  \typeout{******************************************}
  \typeout{** Please run "bibtex \jobname" to optain}
  \typeout{** the bibliography and then re-run LaTeX}
  \typeout{** twice to fix the references!}
  \typeout{******************************************}
  \typeout{}
 }

\end{document}